\documentclass[preprint,showpacs,byrevtex,groupedaddress,onecolumn,pra,11pt]{revtex4}
\usepackage{amsfonts}
\usepackage{amsmath}
\usepackage{amssymb}
\usepackage{graphicx}
\usepackage{dcolumn}
\usepackage{bm}

\begin{document}

\title{Time-dependent localized Hartree-Fock density-functional linear response
approach for photoionization of atomic excited states}
\author{Zhongyuan Zhou}
\author{Shih-I Chu}
\affiliation{Department of Chemistry, University of Kansas, Lawrence, KS 66045}

\begin{abstract}
We present a time-dependent localized Hartree-Fock density-functional linear
response approach for the treatment of photoionization of atomic systems.
This approach employs a spin-dependent localized Hartree-Fock (SLHF)
exchange potential to calculate electron orbitals and kernel functions, and
thus can be used to study the photoionization from atomic excited states. We
have applied the approach to the calculation of photoionization cross
sections of Ne ground state. The results are in agreement with available
experimental data and have comparable accuracies with other \textit{ab initio%
} theoretical results. We have also extended the approach to explore the
photoionization from Ne excited states and obtained some new results for the
photoionization from outer-shell and inner-shell excited states.
\end{abstract}

\date{\today }
\maketitle

\section{Introduction}

Making use of local potentials and independent-particle response functions,
density functional theory (DFT) combined with linear response approximation
(LRA) \cite{Hohenberg64,Kohn65} has been successfully applied to study
dynamic processes, such as photoabsorption \cite%
{Zangwill80,Levine84,Levine87,Tong97,Stener95,Choe93,Stener00,Stener01} and
dynamic polarizability \cite%
{Stott80,Mahan80,Gisbergen95,Gisbergen96,Jamorski96}, of atomic and
molecular systems. The most attractive features of such an approach are its
satisfactory accuracy and computational simplicity and efficiency. However,
since the conventional DFT using traditional exchange-correlation (XC)
potentials obtained from uniform electron gas, such as local density
approximation (LDA) \cite{Dreizler90,Parr89} and generalized gradient
approximation (GGA) \cite{Parr89,Becke88,Perdew86}, is a ground-state
approach, the conventional DFT-LRA approach can only be used to investigate
the dynamic processes associated with the ground state of a system. Even so,
the counterpart of the DFT-LRA approach for the excited states has not yet
been reported.

The difficulty encountered in the extension of the conventional DFT-LRA
approach to the excited states stems from the XC potential used to
characterize the excited states. A qualified XC potential for the excited
states is required to be symmetry-dependent and self-interaction free and
have a correct long-range behavior. The symmetry (such as electronic
configuration, electron orbital angular momentum, and electron spin) of the
XC potential is used to distinguish a state from the others, the
self-interaction-free property of the XC potential is used to make the
calculation of electron orbital energy accurate, and the correct asymptotic
behavior of the XC potential is used to guarantee the Rydberg virtual
orbitals which play a key role in autoionization resonances \cite{Stener95}.

Recently, a localized Hartree-Fock (LHF) density-functional approach has
been proposed and successfully applied to the ground-state calculation of
atomic and molecular systems \cite{a15}. In this approach, the LHF exchange
potential is self-interaction free and exhibits the correct long-range
behavior. It only needs occupied orbitals and depends on the orbital
symmetry of the state. More recently, a spin-dependent localized
Hartree-Fock (SLHF) density-functional approach has been developed for the
excited-state calculation of atomic and molecular systems \cite{zhou2005}.
This approach together with Slater's diagonal sum rule \cite{Slater60} have
been successfully used to calculate the energies of multiply excited states
of valence electrons of atomic systems \cite{zhou2005} and the energies of
inner-shell excited states of closed-shell \cite{zhou07PRA}\ and open-shell 
\cite{zhou07JPB} atomic systems.

In this paper, we present a time-dependent localized Hartree-Fock
density-functional linear response approach for the treatment of
photoionization from atomic excited states by combining the SLHF DFT with
LRA. In this approach, the SLHF exchange potential is employed to calculate
both the Kohn--Sham (KS) electron orbitals and kernel functions. This is
different from the approach used in Ref. \cite{Stener01} where the LHF
potential was only used to calculate the electron orbitals. We have applied
this approach to the calculation of photoionization cross sections (PICS) of
Ne ground state and extended it to the computation of PICS of Ne excited
states.

\section{Theoretical Methodology}

\subsection{Linear response approximation of photoionization}

Suppose that an atomic system is in a time-dependent external field along $z$
axis $\mathcal{E}\left( t\right) =\mathcal{E}_{0}e^{-i\omega t}$, where $%
\mathcal{E}_{0}$ is the amplitude and $\omega $ the frequency. In dipole
approximation, the interaction potential of an electron and the external
field is $\phi ^{ext}\left( \mathbf{r},t\right) =\phi ^{ext}\left( \mathbf{r}%
,\omega \right) e^{-i\omega t}$, where $\phi ^{ext}\left( \mathbf{r},\omega
\right) =z\mathcal{E}_{0}$. Presence of the external field will induce a
perturbation to the system and produce a redistribution of electron density 
\cite{Zangwill80}. In the case of weak field considered here, the dynamic
response of the system has the same time dependence $e^{-i\omega t}$ as the
field \cite{Choe93} and thus can be described by the LRA \cite{Kubo57}. In
frequency domain, total perturbing potential that an electron experiences
can be expressed, in spin-dependent DFT framework \cite%
{Tong97,Stener95,Choe93,Zangwill80}, as 
\begin{equation}
\phi _{\sigma }^{SCF}\left( \mathbf{r},\omega \right) =\phi ^{ext}\left( 
\mathbf{r},\omega \right) +\sum_{\sigma ^{\prime }}\int K_{\sigma \sigma
^{\prime }}\left( \mathbf{r},\mathbf{r}^{\prime }\right) \delta \rho
_{\sigma ^{\prime }}\left( \mathbf{r}^{\prime },\omega \right) d\mathbf{r}%
^{\prime },  \label{1-6}
\end{equation}%
where, the second term on the right-hand side (RHS) is the field-induced
potential, $\sigma $ is the electron spin ($\sigma =$ $\uparrow $ for
spin-up or $\sigma =$ $\downarrow $ for spin-down), $\delta \rho _{\sigma
}\left( \mathbf{r},\omega \right) $ is the field-induced electron density
which is the deviation of the perturbed electron density from the
unperturbed electron density $\rho _{\sigma }\left( \mathbf{r}\right) $, and 
$K_{\sigma \sigma ^{\prime }}\left( \mathbf{r},\mathbf{r}^{\prime }\right) $
is the kernel function%
\begin{equation}
K_{\sigma \sigma ^{\prime }}\left( \mathbf{r},\mathbf{r}^{\prime }\right) =%
\frac{1}{\left\vert \mathbf{r}-\mathbf{r}^{\prime }\right\vert }+\frac{%
\delta V_{xc\sigma }\left( \mathbf{r}\right) }{\delta \rho _{\sigma ^{\prime
}}\left( \mathbf{r}^{\prime }\right) }.  \label{1-7}
\end{equation}%
Here, the first and second terms on the RHS represent the field-induced
changes of Hartree potential and XC potential $V_{xc\sigma }\left( \mathbf{r}%
\right) $, respectively.

The field-induced electron density $\delta \rho _{\sigma }\left( \mathbf{r}%
,\omega \right) $ is related to the total perturbing potential $\phi
_{\sigma }^{SCF}$ by 
\begin{equation}
\delta \rho _{\sigma }\left( \mathbf{r},\omega \right) =\int \chi _{\sigma
}\left( \mathbf{r},\mathbf{r}^{\prime },\omega \right) \phi _{\sigma
}^{SCF}\left( \mathbf{r}^{\prime },\omega \right) d\mathbf{r}^{\prime },
\label{1-8}
\end{equation}%
where, $\chi _{\sigma }$ is the complex susceptibility given by%
\begin{equation}
\chi _{\sigma }\left( \mathbf{r},\mathbf{r}^{\prime },\omega \right)
=\sum_{i}^{occ}\sum_{j}^{all}\left[ \frac{\gamma _{i\sigma }\left( \mathbf{r}%
,\mathbf{r}^{\prime }\right) \gamma _{j\sigma }^{\ast }\left( \mathbf{r},%
\mathbf{r}^{\prime }\right) }{\hbar \omega -\left( \varepsilon _{j\sigma
}-\varepsilon _{i\sigma }\right) +i\epsilon }-\frac{\gamma _{i\sigma }^{\ast
}\left( \mathbf{r},\mathbf{r}^{\prime }\right) \gamma _{j\sigma }\left( 
\mathbf{r},\mathbf{r}^{\prime }\right) }{\hbar \omega +\left( \varepsilon
_{j\sigma }-\varepsilon _{i\sigma }\right) +i\epsilon }\right] .  \label{1-9}
\end{equation}%
Here, $\gamma _{i\sigma }\left( \mathbf{r},\mathbf{r}^{\prime }\right)
=\varphi _{i\sigma }^{\ast }\left( \mathbf{r}\right) \varphi _{i\sigma
}\left( \mathbf{r}^{\prime }\right) $, $\varphi _{i\sigma }\left( \mathbf{r}%
\right) $ is the $i$th electron spin-orbital, $\epsilon $ is a positive
infinitesimal, and the notations $all$ and $occ$ represent that the sums run
over all and occupied electron orbitals, respectively.

From Eqs. (\ref{1-6}) and (\ref{1-8}), $\phi _{\sigma }^{SCF}\left( \mathbf{r%
},\omega \right) $ and $\delta \rho _{\sigma }\left( \mathbf{r},\omega
\right) $ have to be calculated in a self-consistent field procedure.
Applying Eq. (\ref{1-6}) to Eq. (\ref{1-8}) and introducing a kernel function%
\begin{equation}
N_{\sigma \sigma ^{\prime }}\left( \mathbf{r},\mathbf{r}^{\prime },\omega
\right) =\int \chi _{\sigma }\left( \mathbf{r},\mathbf{r}^{\prime \prime
},\omega \right) K_{\sigma \sigma ^{\prime }}\left( \mathbf{r}^{\prime
\prime },\mathbf{r}^{\prime }\right) d\mathbf{r}^{\prime \prime },
\label{1-13-2}
\end{equation}%
one obtains an equation for $\delta \rho _{\sigma }\left( \mathbf{r},\omega
\right) $ 
\begin{equation}
\sum_{\sigma ^{\prime }}\int \left[ \delta _{\sigma \sigma ^{\prime }}\delta
\left( \mathbf{r}-\mathbf{r}^{\prime }\right) -N_{\sigma \sigma ^{\prime
}}\left( \mathbf{r},\mathbf{r}^{\prime },\omega \right) \right] \delta \rho
_{\sigma ^{\prime }}\left( \mathbf{r}^{\prime },\omega \right) d\mathbf{r}%
^{\prime }=\int \chi _{\sigma }\left( \mathbf{r},\mathbf{r}^{\prime },\omega
\right) \phi ^{ext}\left( \mathbf{r}^{\prime },\omega \right) d\mathbf{r}%
^{\prime }.  \label{1-13-3}
\end{equation}

The polarizability $\alpha \left( \omega \right) $ is the ratio of the
induced dipole moment to the external field strength%
\begin{eqnarray}
\alpha \left( \omega \right) &=&-\frac{1}{\mathcal{E}_{0}}\sum_{\sigma }\int
z\delta \rho _{\sigma }\left( \mathbf{r},\omega \right) d\mathbf{r}  \notag
\\
&=&-\frac{1}{\mathcal{E}_{0}^{2}}\sum_{\sigma }\int \int \phi ^{ext}\left( 
\mathbf{r},\omega \right) \chi _{\sigma }\left( \mathbf{r},\mathbf{r}%
^{\prime },\omega \right) \phi _{\sigma }^{SCF}\left( \mathbf{r}^{\prime
},\omega \right) d\mathbf{r}^{\prime }d\mathbf{r}.  \label{8}
\end{eqnarray}

The PICS $\sigma \left( \omega \right) $ is calculated by%
\begin{equation}
\sigma \left( \omega \right) =\frac{4\pi \omega }{c}\text{Im }\alpha \left(
\omega \right) ,  \label{12}
\end{equation}%
where, $c$ is the speed of light.

The LRA is usually referred to as a time-dependent method since it takes the
time-dependent field-induced electron density into account. In contrast, if
the field-induced electron density is neglected in Eq. (\ref{1-6}), one has $%
\phi _{\sigma }^{SCF}=\phi _{\sigma }^{ext}$. In this case, the cross
section Eq. (\ref{8}) is reduced to the result of independent-particle
approximation, which is referred to as a time-independent method.

\subsection{Electron spin-orbitals}

To calculate the occupied electron spin-orbitals we use the SLHF
density-functional approach \cite{zhou2005,zhou07PRA,zhou07JPB}. In this
approach, the electron spin-orbital $\varphi _{i\sigma }\left( \mathbf{r}%
\right) $ and orbital energy $\varepsilon _{i\sigma }$ are calculated from
the KS equation%
\begin{equation}
\left[ -\frac{1}{2}\nabla ^{2}+V_{\sigma }^{\text{eff}}\left( \mathbf{r}%
\right) \right] \varphi _{i\sigma }\left( \mathbf{r}\right) =\varepsilon
_{i\sigma }\varphi _{i\sigma }\left( \mathbf{r}\right) ,  \label{e11}
\end{equation}%
where, $V_{\sigma }^{\text{eff}}$ is the local effective potential given by%
\begin{equation}
V_{\sigma }^{\text{eff}}\left( \mathbf{r}\right) =-\frac{Z}{r}+\sum_{\sigma
}\int \frac{\rho _{\sigma }\left( \mathbf{r}^{\prime }\right) }{\left\vert 
\mathbf{r}-\mathbf{r}^{\prime }\right\vert }d\mathbf{r}^{\prime
}+V_{xc\sigma }\left( \mathbf{r}\right) ,  \label{e11-2}
\end{equation}%
In Eq. (\ref{e11-2}), the spin-dependent electron density is calculated by $%
\rho _{\sigma }\left( \mathbf{r}\right) =\sum_{i=1}^{N_{\sigma }}\left\vert
\varphi _{i\sigma }\left( \mathbf{r}\right) \right\vert ^{2}$, where $%
N_{\sigma }$ is the number of electron with spin $\sigma $. On the RHS of
Eq. (\ref{e11-2}), the first term is Coulomb potential, the second term is
Hartree potential, and the third term $V_{xc\sigma }\left( \mathbf{r}\right) 
$ is the XC potential. The XC potential can be decomposed into exchange
potential $V_{x\sigma }\left( \mathbf{r}\right) $ and correlation potential $%
V_{c\sigma }\left( \mathbf{r}\right) $. The SLHF exchange potential $%
V_{x\sigma }^{\text{SLHF}}(\mathbf{r})$ is given by \cite%
{zhou2005,zhou07PRA,zhou07JPB,Sala01}%
\begin{eqnarray}
V_{x\sigma }^{\text{SLHF}}(\mathbf{r}) &=&-\frac{1}{\rho _{\sigma }(\mathbf{r%
})}\int \frac{\gamma _{\sigma }\left( \mathbf{r},\mathbf{r}^{\prime }\right)
\gamma _{\sigma }\left( \mathbf{r}^{\prime },\mathbf{r}\right) }{\left\vert 
\mathbf{r}-\mathbf{r}^{\prime }\right\vert }d\mathbf{r}^{\prime }  \notag \\
&&+\frac{1}{\rho _{\sigma }(\mathbf{r})}\int \gamma _{\sigma }\left( \mathbf{%
r},\mathbf{r}^{\prime }\right) V_{x\sigma }^{\text{SLHF}}(\mathbf{r}^{\prime
})\gamma _{\sigma }\left( \mathbf{r}^{\prime },\mathbf{r}\right) d\mathbf{r}%
^{\prime }  \notag \\
&&+\frac{1}{\rho _{\sigma }(\mathbf{r})}\int \int \dfrac{\gamma _{\sigma
}\left( \mathbf{r},\mathbf{r}^{\prime }\right) \gamma _{\sigma }\left( 
\mathbf{r}^{\prime },\mathbf{r}^{\prime \prime }\right) \gamma _{\sigma
}\left( \mathbf{r}^{\prime \prime },\mathbf{r}\right) }{\left\vert \mathbf{r}%
^{\prime }-\mathbf{r}^{\prime \prime }\right\vert }d\mathbf{r}^{\prime }d%
\mathbf{r}^{\prime \prime },  \label{e109}
\end{eqnarray}%
where, $\gamma _{\sigma }\left( \mathbf{r},\mathbf{r}^{\prime }\right)
=\sum_{i=1}^{N_{\sigma }}\gamma _{i\sigma }\left( \mathbf{r},\mathbf{r}%
^{\prime }\right) $. On the RHS of Eq. (\ref{e109}), the first term is
Slater potential \cite{Slater60} and the second and third terms are the
corrections to the Slater potential. The $V_{x\sigma }^{\text{SLHF}}(\mathbf{%
r})$ behaves asymptotically as Slater potential and thus approaches to the
correct $-1/r$ at long range \cite{Sala01}. As for the correlation effect,
the second-order gradient correlation potential and energy functional
proposed by Lee, Yang, and Parr (LYP) \cite{Lee88} can provide an excellent
correlation energy for atomic systems and will be incorporated into the
calculation to estimate the correlation effect in this work.

\subsection{Green function}

In Eq. (\ref{1-9}) the sum over $j$ needs all the (occupied and unoccupied
bound and continuum) electron orbitals. This makes it extremely difficult to
accurately calculate $\chi _{\sigma }\left( \mathbf{r},\mathbf{r}^{\prime
},\omega \right) $ directly from Eq. (\ref{1-9}). To circumvent this
difficulty, a Green function associated with the KS equation has been
introduced to calculate $\chi _{\sigma }$ \cite{Tong97,Choe93,Zangwill80}.
The Green function $G_{\sigma }\left( \mathbf{r},\mathbf{r}^{\prime
},E\right) $ is calculated by%
\begin{equation}
\left[ E+\frac{1}{2}\nabla ^{2}-V_{\sigma }^{\text{eff}}\left( \mathbf{r}%
\right) \right] G_{\sigma }\left( \mathbf{r},\mathbf{r}^{\prime },E\right)
=\delta \left( \mathbf{r}-\mathbf{r}^{\prime }\right) ,  \label{31}
\end{equation}%
under appropriate boundary conditions. The Green function can be expanded in
terms of a complete set of KS electron spin-orbitals as \cite{Zangwill80}%
\begin{equation}
G_{\sigma }\left( \mathbf{r},\mathbf{r}^{\prime },E\right) =\sum_{j}^{all}%
\frac{\gamma _{j\sigma }^{\ast }\left( \mathbf{r,r}^{\prime }\right) }{%
E-\varepsilon _{j\sigma }\pm i\epsilon }.  \label{32}
\end{equation}%
Applying Eq. (\ref{32}) to Eq. (\ref{1-9}) one has%
\begin{equation}
\chi _{\sigma }\left( \mathbf{r},\mathbf{r}^{\prime },\omega \right)
=\sum_{i}^{occ}\left[ \gamma _{i\sigma }\left( \mathbf{r},\mathbf{r}^{\prime
}\right) G_{\sigma }\left( \mathbf{r},\mathbf{r}^{\prime },\varepsilon
_{i\sigma }+\hbar \omega \right) +\gamma _{i\sigma }^{\ast }\left( \mathbf{r}%
,\mathbf{r}^{\prime }\right) G_{\sigma }^{\ast }\left( \mathbf{r},\mathbf{r}%
^{\prime },\varepsilon _{i\sigma }-\hbar \omega \right) \right] .  \label{34}
\end{equation}%
Thus with assistance of the Green function, only the occupied orbitals are
needed to compute $\chi _{\sigma }$.

\section{Computational details}

\subsection{Electron spin-orbitals}

The electron spin-orbitals of an atomic system can be calculated by using
the procedure previously developed in \cite{zhou2005,zhou07JPB}. In
spherical coordinates, the electron spin-orbital $\varphi _{i\sigma }\left( 
\mathbf{r}\right) $ is expressed as a product of a radial spin-orbital $%
R_{nl\sigma }(r)$ and a spherical harmonic $Y_{lm}(\theta ,\phi )$%
\begin{equation}
\varphi _{i\sigma }\left( \mathbf{r}\right) =\frac{R_{nl\sigma }(r)}{r}%
Y_{lm}(\theta ,\phi ),  \label{e204}
\end{equation}%
where, $n$ is the principal quantum number, $l$ is the orbital angular
momentum quantum number, $m$ is the azimuthal quantum number, and $i$ is a
set of quantum numbers except the spin $\sigma $. The radial spin-orbital $%
R_{nl\sigma }(r)$ is calculated from the radial KS equation \cite%
{zhou2005,zhou07JPB} 
\begin{equation}
\left[ -\frac{1}{2}\frac{d^{2}}{dr^{2}}+\frac{l(l+1)}{2r^{2}}+v_{\sigma }^{%
\text{eff}}(r)\right] R_{nl\sigma }=\varepsilon _{nl\sigma }R_{nl\sigma },
\label{e205}
\end{equation}%
where, $v_{\sigma }^{\text{eff}}(r)$ is the radial effective potential \cite%
{zhou2005}. To obtain high-precision electron spin-orbital and orbital
energy, we use generalized pseudospectral (GPS) method \cite{Wang94} to
discretize the radial KS equation (\ref{e205}). The GPS method associated
with an appropriate mapping technique can overcome difficulties due to
singularity at $r=0$ and long-tail at large $r$ of the Coulomb interaction.
It allows for nonuniform and optimal spatial discretization with\ the use of
only a modest number of grid points. It has been shown that the GPS method
is a very effective and efficient numerical algorithm for the high-precision
solution of KS equation \cite{zhou2005,zhou07JPB,Chu01a,Chu2005}.

\subsection{Green function}

In spherical coordinates, the Green function can be expanded in terms of
partial waves as%
\begin{equation}
G_{\sigma }\left( \mathbf{r},\mathbf{r}^{\prime },E\right)
=\sum_{LM}Y_{LM}^{\ast }\left( \theta ,\varphi \right) \mathcal{G}_{L\sigma
}\left( r,r^{\prime },E\right) Y_{LM}\left( \theta ^{\prime },\varphi
^{\prime }\right) ,  \label{35}
\end{equation}%
where, $\mathcal{G}_{L\sigma }\left( r,r^{\prime },E\right) $ is the radial
Green function, which, with the appropriate boundary conditions, is
determined by an inhomogeneous equation \cite{Zangwill80,Stott80}%
\begin{equation}
\left[ E+\dfrac{1}{2}\dfrac{1}{r^{2}}\dfrac{\partial }{\partial r}r^{2}%
\dfrac{\partial }{\partial r}-\dfrac{1}{2}\dfrac{L\left( L+1\right) }{r^{2}}%
-v_{\sigma }^{\text{eff}}\left( r\right) \right] \mathcal{G}_{L\sigma
}\left( r,r^{\prime },E\right) =\dfrac{\delta \left( r-r^{\prime }\right) }{%
r^{2}}.  \label{36}
\end{equation}

Alternatively, the radial Green function can also be constructed from the
solutions of a homogeneous equation%
\begin{equation}
\left[ -\dfrac{d^{2}}{dr^{2}}+\dfrac{L\left( L+1\right) }{r^{2}}+2v_{\sigma
}^{\text{eff}}\left( r\right) -k^{2}\right] u_{Lk\sigma }\left( r\right) =0,
\label{36-0-1}
\end{equation}%
where, $k=\sqrt{2E}$. If $\phi _{Lk\sigma }\left( r\right) $ is the solution
of Eq. (\ref{36-0-1}) being regular at the origin and $\psi _{Lk\sigma
}\left( r\right) $ the solution behaving asymptotically as $rh_{L}^{\left(
1\right) }(kr)$ (where $h_{L}^{\left( 1\right) }$ is the spherical Hankel
function of the first kind), the radial Green function can be calculated by 
\cite{Stott80}%
\begin{equation}
\mathcal{G}_{L\sigma }\left( r,r^{\prime },E\right) =\frac{2}{W}\frac{\phi
_{Lk\sigma }\left( r_{<}\right) \psi _{Lk\sigma }\left( r_{>}\right) }{%
rr^{\prime }},  \label{37}
\end{equation}%
where, $r_{<}$ $\left( r_{>}\right) $ refers to the smaller (larger) of $r$
and $r^{\prime }$ and $W=\phi _{Lk\sigma }\psi _{Lk\sigma }^{\prime }-\phi
_{Lk\sigma }^{\prime }\psi _{Lk\sigma }$ is the Wronskian of $\phi
_{Lk\sigma }$ and $\psi _{Lk\sigma }$.

\subsection{Absorber}

In principle, the boundary conditions at $r\rightarrow \infty $ is required
to calculate $\psi _{Lk\sigma }\left( r\right) $. In reality, the boundary
can not be set at $r\rightarrow \infty $. No matter how far the boundary is,
as long as it is located at finite distance, the out-going wavefunction $%
\psi _{Lk\sigma }\left( r\right) $ with $E>0$ may reflect on the boundary,
making the PICS oscillating artificially. To remove the reflection we
introduce an absorber for each out-going wavefunction. The absorber is
characterized by an absorptive potential with a linear dependence of the
radial coordinate%
\begin{equation}
U(r)=\left\{ 
\begin{array}{ll}
0, & 0\leq r<r_{a} \\ 
-U_{0}\dfrac{r-r_{a}}{r_{\max }-r_{a}}, & r_{a}\leq r\leq r_{\max }%
\end{array}%
\right. ,  \label{f-11}
\end{equation}%
where, $r_{\max }$ is the radial coordinate of the boundary, $U_{0}$ and $%
r_{a}$ are two parameters representing the strength and starting position of
the absorber, respectively. Obviously, $r_{\max }-r_{a}$ represents the
width of the absorber. Similar absorbers have been used in the wavepacket
method of molecular collisions \cite{Neuhasuer89,Child91} and
photoionization of molecules and atomic clusters recently \cite%
{Nakatsukasa01}. It is shown that the PICS are insensitive to the absorber
parameters.

When taking the absorptive potential into account the out-going wavefunction
is calculated from an equation obtained by replacing $v_{\sigma }^{\text{eff}%
}\left( r\right) $ with $v_{\sigma }^{\text{eff}}\left( r\right) +iU\left(
r\right) $ in Eq. (\ref{36-0-1}). Since the behavior of an out-going
wavefunction depends on electron spin-orbitals through both $v_{\sigma }^{%
\text{eff}}\left( r\right) $ and $k(E)$ the absorber parameters may be
different for different electron spin-orbitals.

\subsection{Susceptibility and cross sections}

In spherical coordinates, the susceptibility $\chi _{\sigma }\left( \mathbf{r%
},\mathbf{r}^{\prime },\omega \right) $ can also be expanded in the partial
waves \cite{Stott80}%
\begin{equation}
\chi _{\sigma }\left( \mathbf{r},\mathbf{r}^{\prime },\omega \right)
=\sum_{lm}Y_{lm}^{\ast }\left( \theta ,\phi \right) \chi _{l\sigma }\left(
r,r^{\prime },\omega \right) Y_{lm}\left( \theta ^{\prime },\phi ^{\prime
}\right) .  \label{a-1}
\end{equation}%
From Eqs. (\ref{34}), (\ref{e204}), (\ref{35}), and (\ref{a-1}), the partial
wave susceptibility $\chi _{l\sigma }\left( r,r^{\prime },\omega \right) $
is calculated by%
\begin{eqnarray}
\chi _{l\sigma }\left( r,r^{\prime },\omega \right) &=&\frac{1}{4\pi }%
\sum_{n^{\prime }l^{\prime }}\sum_{L}w_{n^{\prime }l^{\prime }\sigma }\frac{%
R_{n^{\prime }l^{\prime }\sigma }(r)}{r}\frac{R_{n^{\prime }l^{\prime
}\sigma }(r^{\prime })}{r^{\prime }}\left\vert \left\langle l0l^{\prime
}0|L0\right\rangle \right\vert ^{2}  \notag \\
&&\times \left[ \mathcal{G}_{L\sigma }\left( r,r^{\prime },\varepsilon
_{n^{\prime }l^{\prime }\sigma }+\hbar \omega \right) +\mathcal{G}_{L\sigma
}^{\ast }\left( r,r^{\prime },\varepsilon _{n^{\prime }l^{\prime }\sigma
}-\hbar \omega \right) \right] .  \label{a-2}
\end{eqnarray}%
Furthermore, expanding $\delta \rho _{\sigma }\left( \mathbf{r},\omega
\right) $ in the partial waves $\delta \rho _{\sigma }\left( \mathbf{r}%
,\omega \right) =\sum_{lm}\delta \rho _{lm\sigma }\left( r,\omega \right)
Y_{lm}\left( \theta ,\phi \right) $, using $z=\sqrt{\frac{4\pi }{3}}%
rY_{10}\left( \theta ,\phi \right) $, and from Eq. (\ref{1-13-3}), we obtain%
\begin{eqnarray}
\delta \rho _{lm\sigma }\left( r,\omega \right) &=&\mathcal{E}_{0}\sqrt{%
\frac{4\pi }{3}}\int \chi _{1\sigma }\left( r,r^{\prime },\omega \right)
r^{\prime 3}dr^{\prime }\delta _{l1}\delta _{m0}  \notag \\
&&+\sum_{\sigma ^{\prime }}\int N_{l\sigma \sigma ^{\prime }}\left(
r,r^{\prime },\omega \right) \delta \rho _{lm\sigma ^{\prime }}\left(
r^{\prime },\omega \right) r^{\prime 2}dr^{\prime },  \label{z-17}
\end{eqnarray}%
where, $N_{l\sigma \sigma ^{\prime }}\left( r,r^{\prime },\omega \right) $
is the partial wave component in the partial wave expansion $N_{\sigma
\sigma ^{\prime }}\left( \mathbf{r},\mathbf{r}^{\prime }\right)
=\sum_{lm}Y_{lm}^{\ast }\left( \theta ,\phi \right) N_{l\sigma \sigma
^{\prime }}\left( r,r^{\prime },\omega \right) Y_{lm}\left( \theta ^{\prime
},\phi ^{\prime }\right) $. Solving Eq. (\ref{z-17}) one obtains the partial
wave component $\delta \rho _{lm\sigma }\left( r,\omega \right) $. From Eq. (%
\ref{8}), the polarizability $\alpha \left( \omega \right) $ is given by%
\begin{equation}
\alpha \left( \omega \right) =-\frac{1}{\mathcal{E}_{0}}\sqrt{\frac{4\pi }{3}%
}\sum_{\sigma }\int \delta \rho _{10\sigma }\left( r,\omega \right) r^{3}dr.
\label{z-18}
\end{equation}%
The PICS $\sigma \left( \omega \right) $ is calculated from Eq. (\ref{12})
by using $\alpha \left( \omega \right) $.

\section{Results and discussion}

\subsection{Photoionization from the ground state of Ne}

To test the approach developed in the preceeding sections we first apply it
to the calculation of PICS from the ground state of Ne. The total PICS is
plotted in FIG. 1, where the solid and dashed lines are the results of
time-dependent spin-dependent localized Hartree-Fock (TDSLHF) method and
time-independent spin-dependent localized Hartree-Fock (TISLHF) method,
respectively. Also plotted in this figure are recent experimental data \cite%
{Chan92,Samson02} for comparison. It shows that the TISLHF PICS are
substantially larger than the experimental results in a broad range of lower
photon energy. While the TDSLHF cross sections are much closer to the
experimental results and show significant improvement in spite of a little
bit underestimate near the ionization threshold (IT) of $2p$ electron. In
the range of photon energy shown in FIG. 1, apart from a small stepwise
enhancement due to the contribution of $2s$ electron photoionization\ at the
IT energy of $2s$ electron $46.405$ $eV$, the TISLHF PICS are structureless
and decrease monotonically above the IT of $2s$ electron. The TDSLHF PICS
however appear a series of sharp resonances in the energy region from $43.0$
to $46.4$ $eV$. In the higher energy region, both TDSLHF and TISLHF results
are getting closer and merge to the experimental results.

The sharp resonances in the PICS are produced by the resonant transitions $%
2s\rightarrow np$ from the ground state $1s^{2}2s^{2}2p^{6}$ to
autoionization states $1s^{2}2s2p^{6}np$ with $n=3$ to $10$. In FIG. 2(a) we
show the details of the resonances in open circles. To quantitatively
analyze the resonances and estimate the autoionization widths we fit the
resonances to the Fano profile \cite{Fano65} 
\begin{equation}
\sigma \left( \omega \right) =\sigma _{0}\left[ \eta ^{2}\frac{\left(
q+\kappa \right) ^{2}}{1+\kappa ^{2}}-\eta ^{2}+1\right] ,  \label{f-20}
\end{equation}%
where, $\kappa =2\left( \omega -E_{r}\right) /\Gamma $, $E_{r}$ is the
resonance position, $q$ is the profile index, $\Gamma $ is the line-width of
the resonance profile, $\eta ^{2}$ is the correlation coefficient, and $%
\sigma _{0}$ is the cross section without correlation. In FIG. 2(a) we show
the fitted resonances in the solid line. The fitted resonance profile
parameters are given in TABLE \ref{T-1} in the rows with TDSLHF. Also listed
in TABLE \ref{T-1} are the available experimental results \cite{Codling67}
and the theoretical results of the time-dependent density functional method
employing exact exchange-only KS potential and
adiabatic-local-density-approximation exchange-correlation kernel with
shifted $2s$ orbital energy (S-EXX/ALDA) \cite{Stener01}, time-dependent
local spin-density approximation with shifted $2s$ orbital energy (S-TDLSDA) 
\cite{Tong97}, time-dependent local density approximation (TDLDA) \cite%
{Stener95}, and R-matrix method (R-matrix) \cite{Schulz96}, for comparison.
It is shown that the line-width $\Gamma $\ and cross section $\sigma _{0}$
of TDSLHF are in overall good agreement with the experimental values and
better than the S-EXX/ALDA, TDLDA and R-matrix results. The line profile
index $q$ of TDSLHF is larger than both the experimental and theoretical
results. Except the S-EXX/ALDA and TDLDA results, the correlation
coefficient $\eta ^{2}$ of TDSLHF is smaller than the others. As for the
resonance position $E_{r}$, all the resonances in TDSLHF PICS are about $2$ $%
eV$ shift to the lower photon energy with respect to the experimental data.
Similar phenomena were also found in \cite{Tong97,Stener01}. This
discrepancy mainly stems from the deviation of the $2s$ electron orbital
energy from the experimental value \cite{Tong97}. For demonstration, we
notice that the $2s$ electron orbital energy of TDSLHF is $-1.707$ a.u.
which is about $2.04$ $eV$ higher than the experimental value $-1.782$ a.u. 
\cite{Sevier79}. To explore influence of the orbital energy to the PICS and
resonance profiles we have also performed a TDSLHF calculation with the $2s$
electron orbital energy being replaced by the experimental value in the
calculation of the Green function. The calculated and fitted PICS in the
resonance energy region are shown in FIG. 2(b) in the open circles and solid
line, respectively. The fitted resonance profile parameters are listed in
TABLE \ref{T-1} in the rows with S-TDSLHF. It is shown that with the $2s$
electron orbital energy replaced by the experimental value the resonances
shift to the higher photon energy about $2$ $eV$ and the calculated
resonance positions $E_{r}$ are in very good agreement with the experimental
results. In addition, all the other resonance profile parameters,
particularly $\Gamma $ and $\eta ^{2}$, are also significantly improved and
much closer to the experimental values.

\subsection{Photoionization from outer-shell excited states of Ne}

As an extension we apply the proposed approach to the computation of PICS of
Ne outer-shell excited states. In FIG. 3, we show, in the solid and dashed
lines, the total PICS of TDSLHF method and TISLHF method for the
photoionization from the outer-shell excited state $1s_{\downarrow
}1s_{\uparrow }2s_{\downarrow }2s_{\uparrow }2p_{\downarrow
}^{3}2p_{\uparrow }^{2}3s_{\uparrow }$ of Ne, respectively. In FIG. 4 we
show the details of autoionization resonances in the total PICS. It is shown
that the total PICS can be divided into five regions A$\sim $E. They are
separated by the steps at ITs in the TISLHF PICS. The first region A starts
from the IT of $3s_{\uparrow }$ electron and ends at the IT of $2p_{\uparrow
}$ electron followed by two series of sharp autoionization resonances in the
TDSLHF PICS. These two series of resonances are produced by the transitions
of a $2p_{\uparrow }$ electron to the higher bound $s_{\uparrow }$ and $%
d_{\uparrow }$ orbitals, respectively, and thus converge to the IT of $%
2p_{\uparrow }$ electron, as shown in FIG. 4(a). Unlike the PICS of the
ground state which increase with photon energy near the IT of $2p$ electron,
the PICS of the excited state decrease with photon energy near the IT of the 
$3s_{\uparrow }$ electron. The second region B starts from the IT of $%
2p_{\uparrow }$ electron and ends at the IT of $2p_{\downarrow }$ electron.
In this region there are also two series of sharp resonances in the PICS, as
shown in FIG. 4(b). They are induced by the transitions of a $2p_{\downarrow
}$ electron to the higher bound $s_{\downarrow }$ and $d_{\downarrow }$
orbitals, respectively, and converge to the IT of $2p_{\downarrow }$
electron. The third region C starts from the IT of $2p_{\downarrow }$
electron and ends at the IT of $2s_{\uparrow }$ electron followed by a
series of sharp resonances produced by the transitions of a $2s_{\uparrow }$
electron to the higher bound $p_{\uparrow }$ orbitals, as shown in FIG.
4(c). Note that one resonance produced by the transition $2s_{\uparrow
}\rightarrow 2p_{\uparrow }$ in this series is located in the lower energy
region A in FIG. 4(a). The fourth region D starts from the IT of $%
2s_{\uparrow }$ electron and ends at the IT of $2s_{\downarrow }$ electron.
This region consists of a series of sharp resonances produced by the
transitions of a $2s_{\downarrow }$ electron to the higher bound $%
p_{\downarrow }$ orbitals as shown in FIG. 4(d). The last region E covers
the energy region from the IT of $2s_{\downarrow }$ electron to the maximum
energy shown in FIG. 3. In this region the PICS are structureless since the
next autoionization resonances occur at very high energy region when a $%
1s_{\uparrow }$ electron is resonantly pumped to the higher bound $%
p_{\uparrow }$ orbitals. We have also performed the calculation of PICS from
other outer-shell excited states of Ne. The total PICS have the similar
structures as those from the excited state $1s_{\downarrow }1s_{\uparrow
}2s_{\downarrow }2s_{\uparrow }2p_{\downarrow }^{3}2p_{\uparrow
}^{2}3s_{\uparrow }$ shown in FIG. 3.

For the photoionization from Ne excited states, particularly for those with
autoionization resonances, both experimental and theoretical results are
scarce. Thus it is hard to make direct and comprehensive comparison of our
results to the experimental and other theoretical results. Through a roughly
qualitative comparison we found that for the photoionization from the
excited state $1s_{\downarrow }1s_{\uparrow }2s_{\downarrow }2s_{\uparrow
}2p_{\downarrow }^{3}2p_{\uparrow }^{2}3p_{\uparrow }$ the results of TDSLHF
method are in the same order of magnitude as the available experimental data 
\cite{Claessens06} and other theoretical results \cite{Duzy80,Chang82-1}.
For example, at the photon energy 3.41 eV, the PICS of TISLHF and TDSLHF
methods are 3.99 Mb and 6.70 Mb, respectively, for the photoionization from
the excited state $1s_{\downarrow }1s_{\uparrow }2s_{\downarrow
}2s_{\uparrow }2p_{\downarrow }^{3}2p_{\uparrow }^{2}3p_{\uparrow }$, while
the experimental results for the photoionization from the excited state $%
1s^{2}2s^{2}2p^{5}3p$ $^{3}D_{3}$ is 2.15 Mb \cite{Claessens06}, the
theoretical results of the central-field approximation for the
photoionization from the excited state $1s^{2}2s^{2}2p^{5}3p$ is about 5.0
Mb \cite{Duzy80}, and the theoretical results of the single-configuration
Hartree-Fock method for the photoionization from the excited state $%
1s^{2}2s^{2}2p^{5}3p$ $^{1}S$ is about 4.5 Mb \cite{Chang82-1}. For the
photoionization from $1s_{\downarrow }1s_{\uparrow }2s_{\downarrow
}2s_{\uparrow }2p_{\downarrow }^{3}2p_{\uparrow }^{2}3s_{\uparrow }$, the
results of TDSLHF method are larger than the available experimental \cite%
{Kau96} and other theoretical results \cite{Hazi77,McCann77,Duzy80,Kau96}.

Interchannel interference has a significant impact on the autoionization
resonances in the photoionization from excited states. Due to the
interference of autoionization channels, the profiles of autoionization
resonances, particularly those produced by the transitions from the same
electron orbital such as those shown in FIG. 4(a) and (b), are no longer the
standard Fano profile. Thus it is impossible to accurately evaluate
autoionization resonance profile parameters by fitting the resonances to
Fano profile. However, we can still estimate the peak position for each
autoionization resonance very well. In TABLE \ref{T-2}, we list the peak
positions of autoionization resonances for the photoionization from Ne
outer-shell excited states $1s_{\downarrow }1s_{\uparrow }2s_{\downarrow
}2s_{\uparrow }2p_{\downarrow }^{3}2p_{\uparrow }^{2}nl_{\uparrow }$ with $%
n=3\sim 4$ and $l=0\sim 2$, where $\{\}=1s_{\downarrow }1s_{\uparrow
}2s_{\downarrow }2s_{\uparrow }2p_{\downarrow }^{3}2p_{\uparrow }^{2}$ is
the abbreviation of the core configuration.

One of the important physical processes during the photoionization is
orbital relaxation \cite{Hazi77}. Due to the orbital relaxation the peak
position of an autoionization resonance is different from the orbital energy
difference between the two unperturbed electron orbitals involved. For
example, for the excited state $1s_{\downarrow }1s_{\uparrow }2s_{\downarrow
}2s_{\uparrow }2p_{\downarrow }^{3}2p_{\uparrow }^{2}3s_{\uparrow }$, the
orbital energy difference between the unperturbed $2p_{\uparrow }$ and $%
2s_{\uparrow }$ electron orbitals is $23.484$ $eV$. While the peak position
of the autoionization resonance produced by the transition $2s_{\uparrow
}\rightarrow 2p_{\uparrow }$ is at $22.994$ $eV$, which is about $0.490$ $eV$
shift toward the lower energy with respect to the unperturbed orbital energy
difference. Our calculation shows that the relaxation effect is larger for
the autoionization resonance produced by the transition from a deeper
inner-shell electron orbital to a lower-lying autoionization electron
orbital.

\subsection{Photoionization from inner-shell excited states of Ne}

As another application\ we extend the proposed approach to the calculation
of PICS of Ne inner-shell excited states. In FIG. 5 we plot the total PICS
of TDSLHF and TISLHF methods in\ the solid and dashed lines, respectively,
for the photoionization from Ne inner-shell excited state $1s_{\uparrow
}1s_{\downarrow }2s_{\uparrow }2p_{\uparrow }^{3}2p_{\downarrow
}^{3}3s_{\downarrow }$. The PICS can be divided into four regions A$\sim $D.
Each region starts from the IT of an electron orbital and ends at the IT of
next lower-energy electron orbital. It is shown that the TISLHF PICS are
structureless apart from a stepwise enhancement at each IT. The TDSLHF PICS,
however, contain several series sharp autoionization resonances, as shown in
FIG. 6.

The region A starts from the IT of $3s_{\downarrow }$ electron and ends at
the IT of $2p_{\downarrow }$ electron followed by two series of
autoionization resonances, as shown in FIG. 6(a). The two series resonances
are produced by the transitions of a $2p_{\downarrow }$ electron to the
higher bound $s_{\downarrow }$ and $d_{\downarrow }$ orbitals and converge
to the IT of $2p_{\downarrow }$ electron. The cross sections near the IT of $%
3s_{\downarrow }$ electron again decrease with photon energy. The region B
starts from the IT of $2p_{\downarrow }$ electron and ends at the IT of $%
2p_{\uparrow }$ electron. In this region there are also two series of
autoionization resonances. They are produced by the transitions of a $%
2p_{\uparrow }$ electron to the higher bound $s_{\uparrow }$ and $%
d_{\uparrow }$ orbitals and converge to the IT of $2p_{\uparrow }$ electron,
as shown in FIG. 6(b). The region C begins with the IT of $2p_{\uparrow }$
electron and ends at the IT of $2s_{\uparrow }$ electron. In this region,
there is a series of autoionization resonances induced by the transitions of
the $2s_{\uparrow }$ electron to the higher bound $p_{\uparrow }$ orbitals,
as shown in FIG. 6(c). The last region D starts from the IT of $2s_{\uparrow
}$ electron and ends at the maximum energy shown in FIG. 5. In this region,
the PICS are structureless since the autoionization resonances will occur at
very high energy.

We have also estimated the peak positions for the autoionization resonances.
In TABLE \ref{T-3}, we list the peak positions of autoionization resonances
for the photoionization from Ne inner-shell excited states $1s_{\uparrow
}1s_{\downarrow }2s_{\uparrow }2p_{\uparrow }^{3}2p_{\downarrow
}^{3}nl_{\downarrow }$ for $n=3\sim 4$ and $l=0\sim 2$. Our calculation
shows again that the relaxation effect is larger for the autoionization
resonance induced by the transition from a deeper inner-shell electron
orbital to a lower-lying autoionization electron orbital.

\section{Conclusion}

In this paper, we propose a time-dependent localized Hartree-Fock
density-functional linear response approach for the treatment of
photoionization of atomic systems. In this approach, the occupied electron
orbitals are calculated by solving the KS equation with SLHF exchange
potential, the complex susceptibility is calculated by using the occupied
electron orbitals and corresponding Green functions, and the PICS are
calculated by using the susceptibility. The relaxation of electron orbitals
is taken into account through the Green functions. To remove the possible
reflection of the wavefunction on the boundary an absorber is imposed for
each out-going wavefunction in the calculation. The absorber is
characterized by an absorptive potential with a linear dependence of
coordinates. Since the SLHF exchange potential has a correct long-range
behavior and can be used to accurately calculate the electron orbitals of
atomic excited states, the proposed approach is suitable for the calculation
of photoionization from excited states of atomic systems. We have applied
this approach to the calculation of PICS of Ne ground state. The results are
in agreement with available experimental and have comparable accuracies with
other \textit{ab initio} theoretical results. We have also extended this
approach to the computation of PICS of Ne excited states. The total PICS can
be divided into several regions. Each region starts from the IT of an
electron orbital and ends at the IT of the next lower-energy electron
orbital. In each region there is one or two series of autoionization
resonances produced by the resonant transitions of an electron from
inner-shell electron orbitals to autoionization electron orbitals with
higher energies. The orbital relaxation effect is larger for the
autoionization resonance produced by the transitions from deeper inner-shell
electron orbitals to lower-lying autoionization electron orbitals.

\begin{acknowledgments}
This work is partially supported by the Chemical Sciences, Geosciences and
Biosciences Division of the Office of Basic Energy Sciences, Office of
Science, U. S. Department of Energy, and by the National Science Foundation.
\end{acknowledgments}

\bibliographystyle{apsrev}
\bibliography{dft2}

\newpage 

\textbf{Figure Captions}

FIG. 1 (Color online) Total photoionization cross sections from the ground
state of Ne. The solid and dashed lines are the results of TDSLHF and TISLHF
methods, respectively. The open and solid circles are the experimental data
of Chan \textit{et al}. \cite{Chan92} and Samson \textit{et al}. \cite%
{Samson02}.

FIG. 2 (Color online) Autoionization resonances in the total photoionization
cross sections near the ionization threshold of $2p$ electron. (a) The
results of TDSLHF method and (b) the results of TDSLHF method with the $2s$
electron orbital energy being replaced by the experimental value. The
resonance peaks $1$ to $8$ are produced by the transitions $2s\rightarrow np$
with $n=3$ to $10$, respectively. The open circles are the numerical results
and the solid lines are the fitted results to the Fano profile.

FIG. 3 (Color online) Total photoionization cross sections of Ne outer-shell
excited state $1s_{\downarrow }1s_{\uparrow }2s_{\downarrow }2s_{\uparrow
}2p_{\downarrow }^{3}2p_{\uparrow }^{2}3s_{\uparrow }$. The solid and dashed
lines are the results of TDSLHF and TISLHF methods, respectively.

FIG. 4 (Color online) Autoionization resonances in the total photoionization
cross sections of Ne outer-shell excited state $1s_{\downarrow }1s_{\uparrow
}2s_{\downarrow }2s_{\uparrow }2p_{\downarrow }^{3}2p_{\uparrow
}^{2}3s_{\uparrow }$. (a) The resonances produced by the transitions $%
2p_{\uparrow }\rightarrow ns_{\uparrow }$ and $n^{\prime }d_{\uparrow }$
with $n\geq 4$ and $n^{\prime }\geq 3$. Note that the resonance produced by
the transition $2s_{\uparrow }\rightarrow 2p_{\uparrow }$ is embodied in
this region. (b) The resonances produced by the transitions $2p_{\downarrow
}\rightarrow ns_{\downarrow }$ and $nd_{\downarrow }$ with $n\geq 3$. (c)
The resonances produced by the transitions $2s_{\uparrow }\rightarrow
np_{\uparrow }$ with $n\geq 2$. Note that the resonance for $n=2$ is located
in the energy region in (a). (d) The resonances produced by the transitions $%
2s_{\downarrow }\rightarrow np_{\downarrow }$ with $n\geq 3$.

FIG. 5 (Color online) Total photoionization cross sections of Ne inner-shell
excited state $1s_{\uparrow }1s_{\downarrow }2s_{\uparrow }2p_{\uparrow
}^{3}2p_{\downarrow }^{3}3s_{\downarrow }$. The solid and dashed lines are
the results of TDSLHF and TISLHF methods, respectively.

FIG. 6 (Color online) Autoionization resonances in the total photoionization
cross sections of Ne inner-shell excited state $1s_{\uparrow }1s_{\downarrow
}2s_{\uparrow }2p_{\uparrow }^{3}2p_{\downarrow }^{3}3s_{\downarrow }$. (a)
The resonances produced by the transitions $2p_{\downarrow }\rightarrow
ns_{\downarrow }$ and $n^{\prime }d_{\downarrow }$ with $n\geq 4$ and $%
n^{\prime }\geq 3$. (b) The resonances produced by the transitions $%
2p_{\uparrow }\rightarrow ns_{\uparrow }$ and $nd_{\uparrow }$ with $n\geq 3$%
. (c) The resonances produced by the transitions $2s_{\uparrow }\rightarrow
np_{\uparrow }$ with $n\geq 3$.

\newpage 

\begingroup\squeezetable

\begin{table}[hptb] \centering%
\caption{Profile parameters of the autoionization resonances produced by 
the transitions from the ground state of Ne to the Rydberg series 
$1s^{2}2s2p^{6}np$ with $n=3$ to $8$.\label{T-1}}%

\begin{tabular}{cclcccccccccc}
\hline\hline
Resonant transition &  & Approach &  & $\sigma _{0}\left( Mb\right) $ &  & $%
\Gamma \left( meV\right) $ &  & $E_{r}\left( eV\right) $ &  & $q$ &  & $\eta
^{2}$ \\ \hline
$2s\rightarrow 3p$ &  & TDSLHF$^{a}$ &  & $8.34$ &  & $12.80$ &  & $43.358$
&  & $-4.03$ &  & $0.551$ \\ 
&  & S-TDSLHF$^{b}$ &  & $8.40$ &  & $15.23$ &  & $45.447$ &  & $-3.40$ &  & 
$0.619$ \\ 
&  & S-EXX/ALDA$^{c}$ &  & $8.09$ &  & $17.9$ &  & $45.438$ &  & $-3.18$ & 
& $0.547$ \\ 
&  & S-TDLSDA$^{d}$ &  & $8.28$ &  & $18.04$ &  & $45.453$ &  & $-2.40$ &  & 
$0.764$ \\ 
&  & TDLDA$^{e}$ &  & $8.18$ &  & $13.90$ &  & $46.253$ &  & $-3.69$ &  & $%
0.514$ \\ 
&  & R-matrix$^{f}$ &  &  &  & $34.9$ &  & $45.534$ &  &  &  &  \\ 
&  & Expt.$^{g}$ &  & $8.6\pm 0.6$ &  & $13\pm 2$ &  & $45.546\pm 0.008$ & 
& $-1.6\pm 0.2$ &  & $0.70\pm 0.07$ \\ 
&  &  &  &  &  &  &  &  &  &  &  &  \\ 
$2s\rightarrow 4p$ &  & TDSLHF$^{a}$ &  & $8.12$ &  & $3.89$ &  & $45.028$ & 
& $-4.35$ &  & $0.543$ \\ 
&  & S-TDSLHF$^{b}$ &  & $8.15$ &  & $4.55$ &  & $47.115$ &  & $-3.74$ &  & $%
0.606$ \\ 
&  & S-EXX/ALDA$^{c}$ &  & $7.89$ &  & $5.5$ &  & $47.093$ &  & $-3.35$ &  & 
$0.528$ \\ 
&  & S-TDLSDA$^{d}$ &  & $8.06$ &  & $5.14$ &  & $47.098$ &  & $-2.62$ &  & $%
0.783$ \\ 
&  & TDLDA$^{e}$ &  & $7.98$ &  & $3.86$ &  & $47.397$ &  & $-3.95$ &  & $%
0.505$ \\ 
&  & R-matrix$^{f}$ &  &  &  & $6.65$ &  & $47.111$ &  &  &  &  \\ 
&  & Expt.$^{g}$ &  & $8.0\pm 0.6$ &  & $4.5\pm 1.5$ &  & $47.121\pm 0.005$
&  & $-1.6\pm 0.3$ &  & $0.70\pm 0.07$ \\ 
&  &  &  &  &  &  &  &  &  &  &  &  \\ 
$2s\rightarrow 5p$ &  & TDSLHF$^{a}$ &  & $8.05$ &  & $1.55$ &  & $45.625$ & 
& $-4.47$ &  & $0.537$ \\ 
&  & S-TDSLHF$^{b}$ &  & $8.06$ &  & $1.81$ &  & $47.713$ &  & $-3.85$ &  & $%
0.600$ \\ 
&  & S-TDLSDA$^{d}$ &  & $7.91$ &  & $2.20$ &  & $47.683$ &  & $-2.72$ &  & $%
0.783$ \\ 
&  & TDLDA$^{e}$ &  & $7.91$ &  & $1.62$ &  & $47.814$ &  & $-4.05$ &  & $%
0.502$ \\ 
&  & R-matrix$^{f}$ &  &  &  & $2.47$ &  & $47.692$ &  &  &  &  \\ 
&  & Expt.$^{g}$ &  & $8.2\pm 0.6$ &  & $2\pm 1$ &  & $47.692\pm 0.005$ &  & 
$-1.6\pm 0.5$ &  & $0.70\pm 0.14$ \\ 
&  &  &  &  &  &  &  &  &  &  &  &  \\ 
$2s\rightarrow 6p$ &  & TDSLHF$^{a}$ &  & $8.00$ &  & $0.89$ &  & $45.901$ & 
& $-4.57$ &  & $0.539$ \\ 
&  & S-TDSLHF$^{b}$ &  & $8.00$ &  & $1.04$ &  & $47.989$ &  & $-3.94$ &  & $%
0.601$ \\ 
&  & R-matrix$^{f}$ &  &  &  & $1.28$ &  & $47.967$ &  &  &  &  \\ 
&  & Expt.$^{g}$ &  &  &  &  &  & $47.967\pm 0.006$ &  &  &  &  \\ 
&  &  &  &  &  &  &  &  &  &  &  &  \\ 
$2s\rightarrow 7p$ &  & TDSLHF$^{a}$ &  & $7.99$ &  & $0.43$ &  & $46.075$ & 
& $-4.99$ &  & $0.555$ \\ 
&  & S-TDSLHF$^{b}$ &  & $8.01$ &  & $0.59$ &  & $48.163$ &  & $-3.88$ &  & $%
0.585$ \\ 
&  & R-matrix$^{f}$ &  &  &  & $0.73$ &  & $48.119$ &  &  &  &  \\ 
&  & Expt.$^{g}$ &  &  &  &  &  & $48.116\pm 0.006$ &  &  &  &  \\ 
&  &  &  &  &  &  &  &  &  &  &  &  \\ 
$2s\rightarrow 8p$ &  & TDSLHF$^{a}$ &  & $7.94$ &  & $0.37$ &  & $46.171$ & 
& $-4.54$ &  & $0.534$ \\ 
&  & S-TDSLHF$^{b}$ &  & $7.96$ &  & $0.42$ &  & $48.258$ &  & $-3.92$ &  & $%
0.605$ \\ 
&  & R-matrix$^{f}$ &  &  &  & $0.46$ &  & $48.211$ &  &  &  &  \\ 
&  & Expt.$^{g}$ &  &  &  &  &  & $48.207\pm 0.006$ &  &  &  &  \\ 
\hline\hline
\end{tabular}

$^{a}$ TDSLHF results. $^{b}$ TDSLHF results with the experimental $2s$
electron orbital energy. $^{c}$ EXX/ALDA results with the experimental $2s$
electron orbital energy \cite{Stener01}. $^{d}$ TDLSDA results with the
experimental $2s$ electron orbital energy \cite{Tong97}. $^{e}$ TDLDA
results \cite{Stener95}. $^{f}$ R-matrix results \cite{Schulz96}. $^{g}$
experimental results \cite{Codling67}.%
\end{table}%
\endgroup

\begingroup\squeezetable

\begin{table}[hptb] \centering%
\caption{Autoionization resonance peak positions (in eV) for the photoionizations from Ne outer-shell excited states $\left \{1s_{\downarrow }1s_{\uparrow
}2s_{\downarrow }2s_{\uparrow }2p_{\downarrow }^{3}2p_{\uparrow }^{2}\right \}
nl_{\uparrow }$ with $n=3\sim 4$ and $l=0\sim 2$.\label{T-2}}%

\begin{tabular}{ccccccccccccc}
\hline\hline
Resonant & \ \ \ \ \ \ \ \  &  & \ \ \ \  & Initial & \ \ \ \  &  & \ \ \ \
\  &  & \ \ \ \  & states & \ \ \ \  &  \\ \cline{3-13}
transition &  & $\{\}^{\ast }$3s$_{\uparrow }$ &  & $\{\}$3p$_{\uparrow }$ & 
& $\{\}$3d$_{\uparrow }$ &  & $\{\}$4s$_{\uparrow }$ &  & $\{\}$4p$%
_{\uparrow }$ &  & $\{\}$4d$_{\uparrow }$ \\ \hline
\multicolumn{1}{r}{2p$_{\uparrow }\rightarrow $3s$_{\uparrow }$} &  & $%
\mathbf{-}$ &  & 23.829 &  & 26.150 &  & 23.527 &  & 24.863 &  & 26.509 \\ 
\multicolumn{1}{r}{4s$_{\uparrow }$} &  & 24.526 &  & 26.738 &  & 29.139 & 
& $\mathbf{-}$ &  & 27.693 &  & 29.554 \\ 
\multicolumn{1}{r}{5s$_{\uparrow }$} &  & 25.383 &  & 27.630 &  & 30.153 & 
& 26.839 &  & 28.581 &  & 30.452 \\ 
\multicolumn{1}{r}{6s$_{\uparrow }$} &  & 25.759 &  & 28.014 &  & 30.575 & 
& 27.206 &  & 28.972 &  & 30.896 \\ 
\multicolumn{1}{r}{7s$_{\uparrow }$} &  & 26.013 &  & 28.221 &  & 30.803 & 
& 27.405 &  & 29.175 &  & 31.119 \\ 
\multicolumn{1}{r}{8s$_{\uparrow }$} &  &  &  & 28.406 &  & 30.928 &  & 
27.549 &  & 29.293 &  & 31.250 \\ 
\multicolumn{1}{r}{9s$_{\uparrow }$} &  &  &  &  &  & 31.043 &  &  &  & 
29.386 &  & 31.334 \\ 
\multicolumn{1}{r}{10s$_{\uparrow }$} &  &  &  &  &  &  &  &  &  &  &  & 
31.416 \\ 
&  &  &  &  &  &  &  &  &  &  &  &  \\ 
\multicolumn{1}{r}{3d$_{\uparrow }$} &  & 24.855 &  & 27.078 &  & 29.565 & 
& 26.371 &  & 28.060 &  & 30.001 \\ 
\multicolumn{1}{r}{4d$_{\uparrow }$} &  & 25.516 &  & 27.769 &  & 30.330 & 
& 26.999 &  & 28.730 &  & 30.670 \\ 
\multicolumn{1}{r}{5d$_{\uparrow }$} &  & 25.826 &  & 28.085 &  & 30.670 & 
& 27.290 &  & 29.046 &  & 30.986 \\ 
\multicolumn{1}{r}{6d$_{\uparrow }$} &  & 26.066 &  & 28.264 &  & 30.850 & 
& 27.453 &  & 29.217 &  & 31.173 \\ 
\multicolumn{1}{r}{7d$_{\uparrow }$} &  &  &  &  &  & 30.961 &  & 27.590 & 
& 29.320 &  & 31.282 \\ 
\multicolumn{1}{r}{8d$_{\uparrow }$} &  &  &  &  &  &  &  &  &  & 29.410 & 
& 31.358 \\ 
&  &  &  &  &  &  &  &  &  &  &  &  \\ 
\multicolumn{1}{r}{2p$_{\downarrow }\rightarrow $3s$_{\downarrow }$} &  & 
29.984 &  & 30.382 &  & 30.656 &  & 30.594 &  & 30.587 &  & 30.643 \\ 
\multicolumn{1}{r}{4s$_{\downarrow }$} &  & 35.446 &  & 36.948 &  & 38.389 & 
& 38.224 &  & 38.302 &  & 38.559 \\ 
\multicolumn{1}{r}{5s$_{\downarrow }$} &  & 36.809 &  & 38.447 &  & 40.406 & 
& 40.463 &  & 40.722 &  & 41.081 \\ 
\multicolumn{1}{r}{6s$_{\downarrow }$} &  &  &  &  &  &  &  &  &  & 42.194 & 
& 42.635 \\ 
&  &  &  &  &  &  &  &  &  &  &  &  \\ 
\multicolumn{1}{r}{3d$_{\downarrow }$} &  & 36.036 &  & 37.378 &  & 38.327 & 
& 38.161 &  & 38.213 &  & 38.412 \\ 
\multicolumn{1}{r}{4d$_{\downarrow }$} &  & 37.018 &  & 38.580 &  & 40.479 & 
& 40.406 &  & 40.697 &  & 41.021 \\ 
\multicolumn{1}{r}{5d$_{\downarrow }$} &  &  &  &  &  &  &  & 41.693 &  & 
41.995 &  & 42.423 \\ 
&  &  &  &  &  &  &  &  &  &  &  &  \\ 
\multicolumn{1}{r}{2s$_{\uparrow }\rightarrow $2p$_{\uparrow }$} &  & 22.994
&  & 23.032 &  & 23.032 &  & 23.048 &  & 23.081 &  & 23.056 \\ 
\multicolumn{1}{r}{3p$_{\uparrow }$} &  & 46.964 &  & 49.187 &  & 51.685 & 
& 48.719 &  & 50.243 &  & 52.129 \\ 
\multicolumn{1}{r}{4p$_{\uparrow }$} &  & 48.510 &  & 50.798 &  & 53.310 & 
& 50.044 &  & 51.767 &  & 53.682 \\ 
\multicolumn{1}{r}{5p$_{\uparrow }$} &  & 49.075 &  & 51.381 &  & 53.960 & 
& 50.572 &  & 52.352 &  & 54.265 \\ 
\multicolumn{1}{r}{6p$_{\uparrow }$} &  & 49.359 &  & 51.659 &  & 54.260 & 
& 50.833 &  & 52.630 &  & 54.577 \\ 
\multicolumn{1}{r}{7p$_{\uparrow }$} &  & 49.623 &  & 51.835 &  & 54.420 & 
& 50.986 &  & 52.782 &  & 54.745 \\ 
\multicolumn{1}{r}{8p$_{\uparrow }$} &  &  &  & 52.036 &  & 54.532 &  & 
51.136 &  & 52.883 &  & 54.845 \\ 
\multicolumn{1}{r}{9p$_{\uparrow }$} &  &  &  &  &  & 54.654 &  &  &  & 
52.978 &  & 54.923 \\ 
\multicolumn{1}{r}{10p$_{\uparrow }$} &  &  &  &  &  &  &  &  &  & 53.098 & 
& 55.013 \\ 
\multicolumn{1}{r}{11p$_{\uparrow }$} &  &  &  &  &  &  &  &  &  &  &  & 
55.114 \\ 
&  &  &  &  &  &  &  &  &  &  &  &  \\ 
\multicolumn{1}{r}{2s$_{\downarrow }\rightarrow $3p$_{\downarrow }$} &  & 
56.790 &  & 57.457 &  & 57.892 &  & 57.786 &  & 57.797 &  & 57.911 \\ 
\multicolumn{1}{r}{4p$_{\downarrow }$} &  & 59.650 &  & 61.242 &  & 62.926 & 
& 62.818 &  & 62.940 &  & 63.218 \\ 
\multicolumn{1}{r}{5p$_{\downarrow }$} &  & 60.817 &  & 62.458 &  & 64.412 & 
& 64.404 &  & 64.747 &  & 65.147 \\ \hline\hline
\end{tabular}

$^{\ast }$ $\{\}=1s_{\downarrow }1s_{\uparrow }2s_{\downarrow }2s_{\uparrow
}2p_{\downarrow }^{3}2p_{\uparrow }^{2}$ is the abbreviation of the core
electron configuration.%
\end{table}%
\endgroup

\begingroup\squeezetable

\begin{table}[hptb] \centering%
\caption{Autoionization resonance peak positions (in eV) for the photoionizations 
from Ne inner-shell excited states $\left \{1s_{\uparrow }1s_{\downarrow
}2s_{\uparrow }2p_{\uparrow }^{3}2p_{\downarrow }^{3}\right \} nl_{\downarrow }
$ with $n=3\sim 4$ and $l=0\sim 2$.\label{T-3}}%

\begin{tabular}{ccccccccccccc}
\hline\hline
Resonant & \ \ \ \ \ \ \ \  &  & \ \ \ \ \  & Initial & \ \ \ \ \  &  & \ \
\ \ \  &  & \ \ \ \ \  & states & \ \ \ \ \  &  \\ \cline{3-13}
transition &  & $\{\}^{\ast }$3s$_{\downarrow }$ & \ \ \ \ \ \ \  & $\{\}$3p$%
_{\downarrow }$ & \ \ \ \ \ \ \  & $\{\}$3d$_{\downarrow }$ & \ \ \ \ \ \ \ 
& $\{\}$4s$_{\downarrow }$ & \ \ \ \ \ \ \  & $\{\}$4p$_{\downarrow }$ & \ \
\ \ \ \ \  & $\{\}$4d$_{\downarrow }$ \\ \hline
\multicolumn{1}{r}{2p$_{\downarrow }\rightarrow $3s$_{\downarrow }$} &  & $%
\mathbf{-}$ &  & 21.655 &  & 23.821 &  & 21.478 &  & 22.681 &  & 24.128 \\ 
\multicolumn{1}{r}{4s$_{\downarrow }$} & \multicolumn{1}{r}{} & 22.188 &  & 
24.648 &  & 26.738 &  & $\mathbf{-}$ &  & 25.356 &  & 27.086 \\ 
\multicolumn{1}{r}{5s$_{\downarrow }$} &  & 23.026 &  & 25.595 &  & 27.742 & 
& 24.626 &  & 26.245 &  & 27.973 \\ 
\multicolumn{1}{r}{6s$_{\downarrow }$} &  & 23.394 &  & 25.998 &  & 28.167 & 
& 24.994 &  & 26.656 &  & 28.403 \\ 
\multicolumn{1}{r}{7s$_{\downarrow }$} &  & 23.644 &  & 26.215 &  & 28.381 & 
& 25.187 &  & 26.866 &  & 28.629 \\ 
\multicolumn{1}{r}{8s$_{\downarrow }$} &  &  &  & 26.411 &  & 28.509 &  & 
25.328 &  & 26.988 &  & 28.756 \\ 
\multicolumn{1}{r}{9s$_{\downarrow }$} &  &  &  &  &  & 28.624 &  & 25.499 & 
& 27.094 &  & 28.841 \\ 
\multicolumn{1}{r}{10s$_{\downarrow }$} &  &  &  &  &  & 28.768 &  &  &  & 
27.224 &  & 28.947 \\ 
&  &  &  &  &  &  &  &  &  &  &  &  \\ 
\multicolumn{1}{r}{3d$_{\downarrow }$} &  & 22.488 &  & 24.923 &  & 27.141 & 
& 24.163 &  & 25.739 &  & 27.508 \\ 
\multicolumn{1}{r}{4d$_{\downarrow }$} &  & 23.143 &  & 25.720 &  & 27.908 & 
& 24.784 &  & 26.398 &  & 28.150 \\ 
\multicolumn{1}{r}{5d$_{\downarrow }$} &  & 23.453 &  & 26.060 &  & 28.248 & 
& 25.073 &  & 26.730 &  & 28.493 \\ 
\multicolumn{1}{r}{6d$_{\downarrow }$} &  & 23.686 &  & 26.254 &  & 28.427 & 
& 25.233 &  & 26.907 &  & 28.678 \\ 
\multicolumn{1}{r}{7d$_{\downarrow }$} &  &  &  & 26.444 &  & 28.542 &  & 
25.369 &  & 27.015 &  & 28.790 \\ 
\multicolumn{1}{r}{8d$_{\downarrow }$} &  &  &  &  &  & 28.654 &  & 25.543 & 
& 27.118 &  & 28.866 \\ 
\multicolumn{1}{r}{9d$_{\downarrow }$} &  &  &  &  &  & 28.801 &  &  &  & 
27.249 &  & 29.048 \\ 
&  &  &  &  &  &  &  &  &  &  &  &  \\ 
\multicolumn{1}{r}{2p$_{\uparrow }\rightarrow $3s$_{\uparrow }$} &  & 30.711
&  & 30.942 &  & 31.198 &  & 31.122 &  & 31.135 &  & 31.201 \\ 
\multicolumn{1}{r}{4s$_{\uparrow }$} &  & 36.393 &  & 37.647 &  & 39.089 & 
& 38.904 &  & 39.002 &  & 39.263 \\ 
\multicolumn{1}{r}{5s$_{\uparrow }$} &  & 37.780 &  & 39.152 &  & 41.141 & 
& 41.176 &  & 41.440 &  & 41.821 \\ 
\multicolumn{1}{r}{6s$_{\uparrow }$} &  &  &  &  &  &  &  &  &  & 42.937 & 
& 43.440 \\ 
&  &  &  &  &  &  &  &  &  &  &  &  \\ 
\multicolumn{1}{r}{3d$_{\uparrow }$} &  & 36.994 &  & 38.088 &  & 39.037 & 
& 38.842 &  & 38.918 &  & 38.108 \\ 
\multicolumn{1}{r}{4d$_{\uparrow }$} &  & 37.987 &  & 39.288 &  & 41.206 & 
& 41.127 &  & 41.421 &  & 41.750 \\ 
\multicolumn{1}{r}{5d$_{\uparrow }$} &  &  &  &  &  & 42.412 &  & 42.390 & 
& 42.725 &  & 43.190 \\ 
&  &  &  &  &  &  &  &  &  &  &  &  \\ 
\multicolumn{1}{r}{2s$_{\uparrow }\rightarrow $3p$_{\uparrow }$} &  & 57.854
&  & 58.344 &  & 58.768 &  & 58.638 &  & 58.676 &  & 58.785 \\ 
\multicolumn{1}{r}{4p$_{\uparrow }$} &  & 60.815 &  & 62.154 &  & 63.862 & 
& 63.729 &  & 63.865 &  & 64.151 \\ 
\multicolumn{1}{r}{5p$_{\uparrow }$} &  & 61.998 &  & 63.375 &  & 65.351 & 
& 65.337 &  & 65.686 &  & 66.107 \\ \hline\hline
\end{tabular}

$^{\ast }$ $\{\}=1s_{\uparrow }1s_{\downarrow }2s_{\uparrow }2p_{\uparrow
}^{3}2p_{\downarrow }^{3}$ is the abbreviation of the core electron
configuration.%
\end{table}%
\endgroup

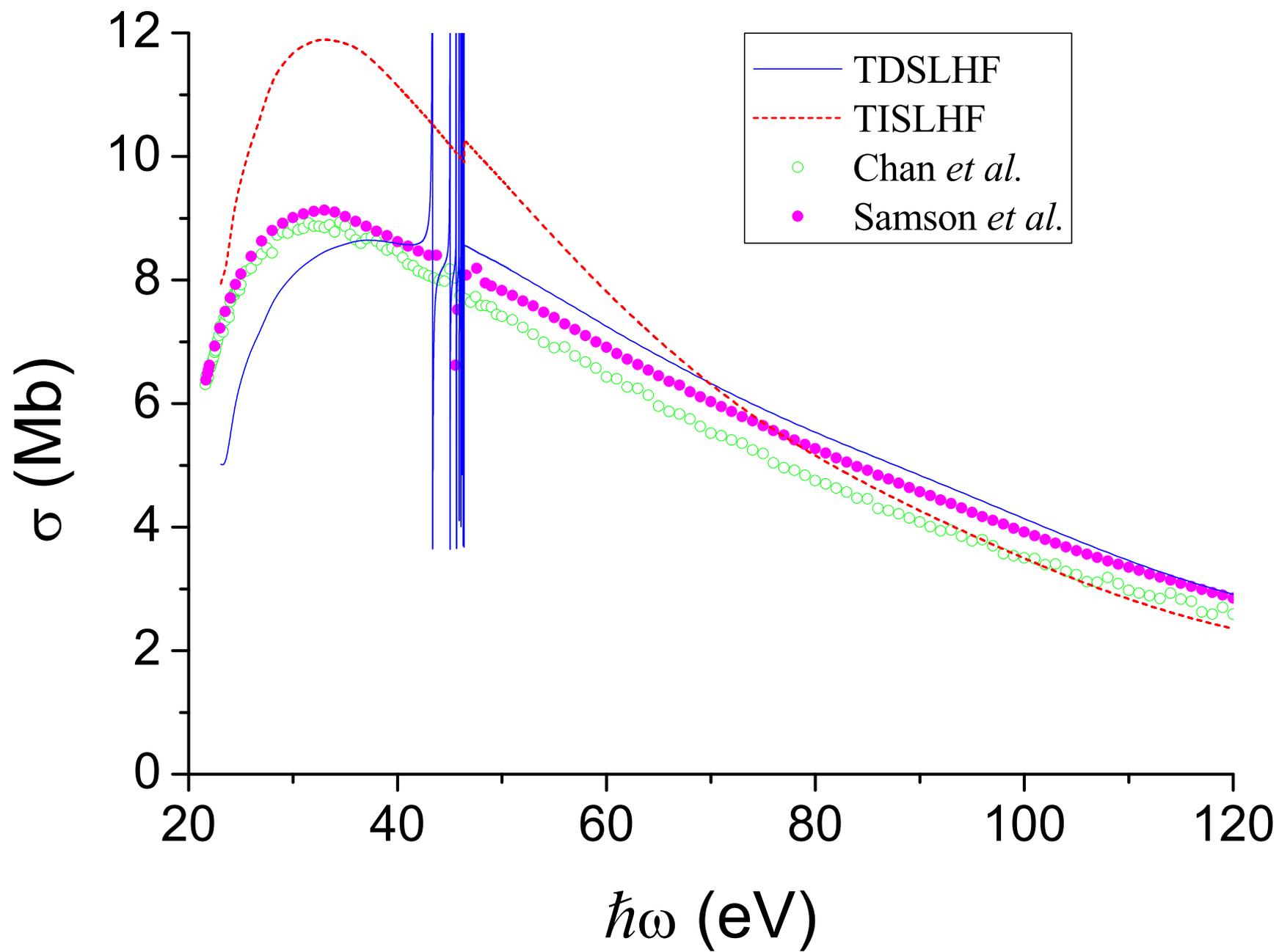
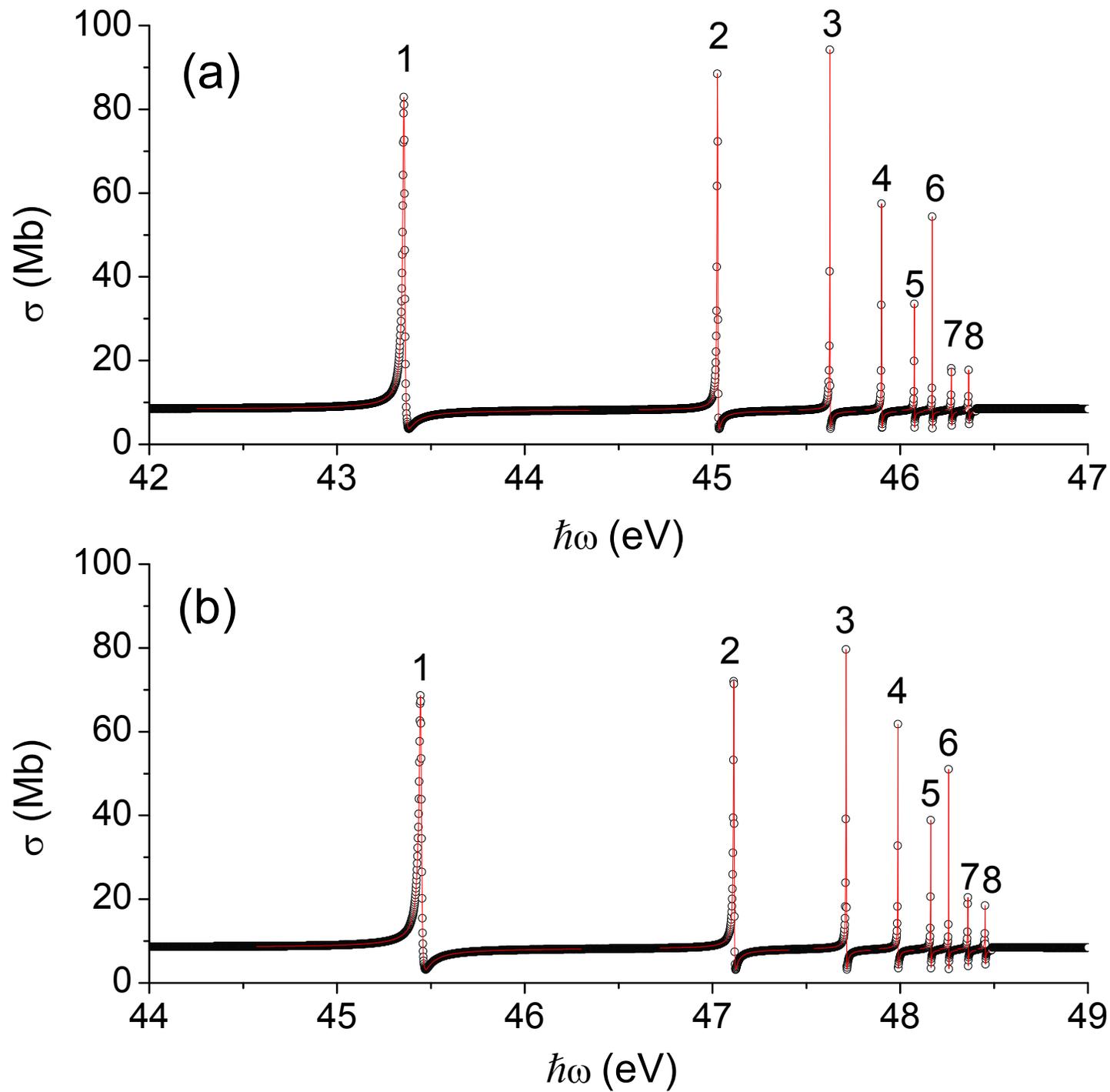
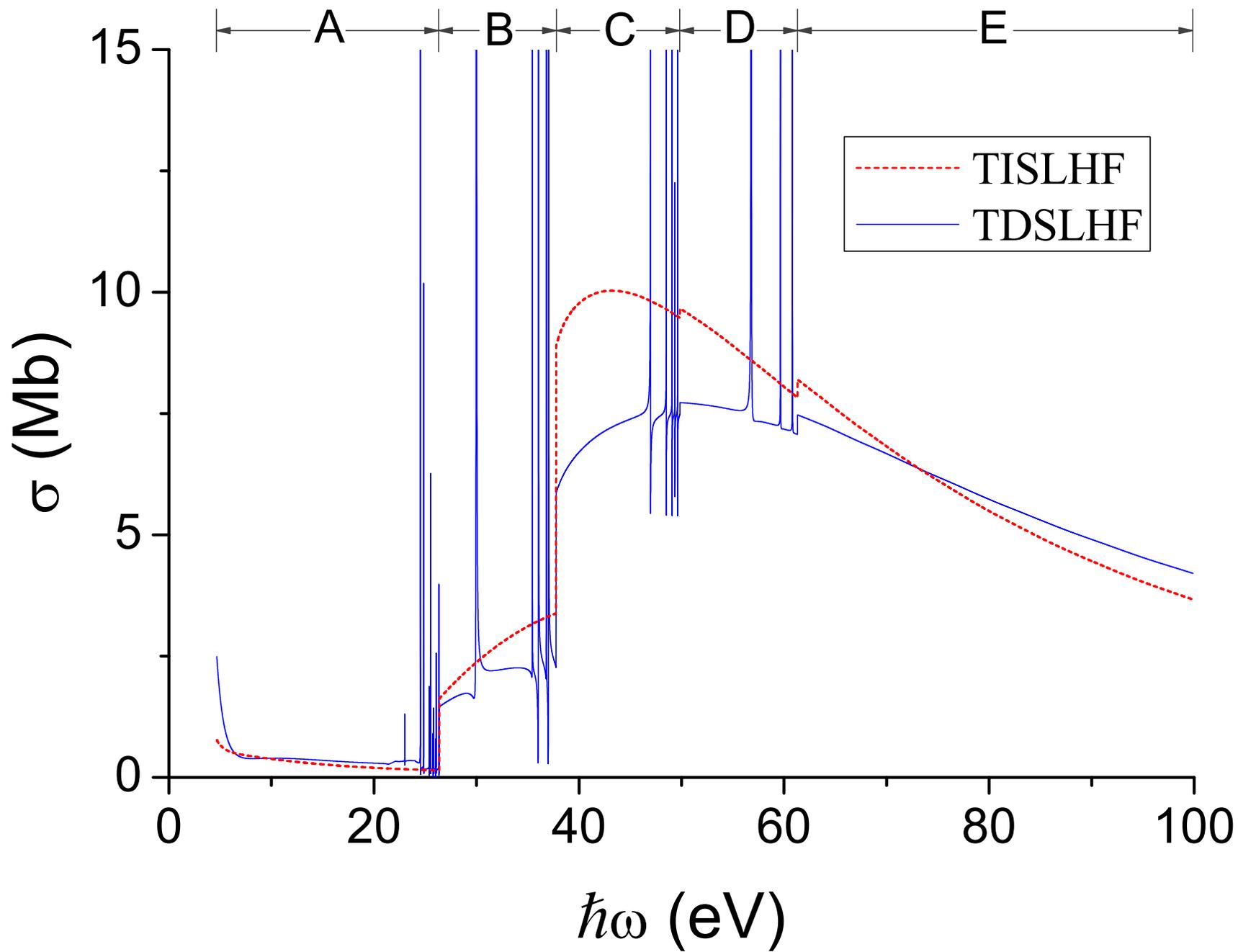
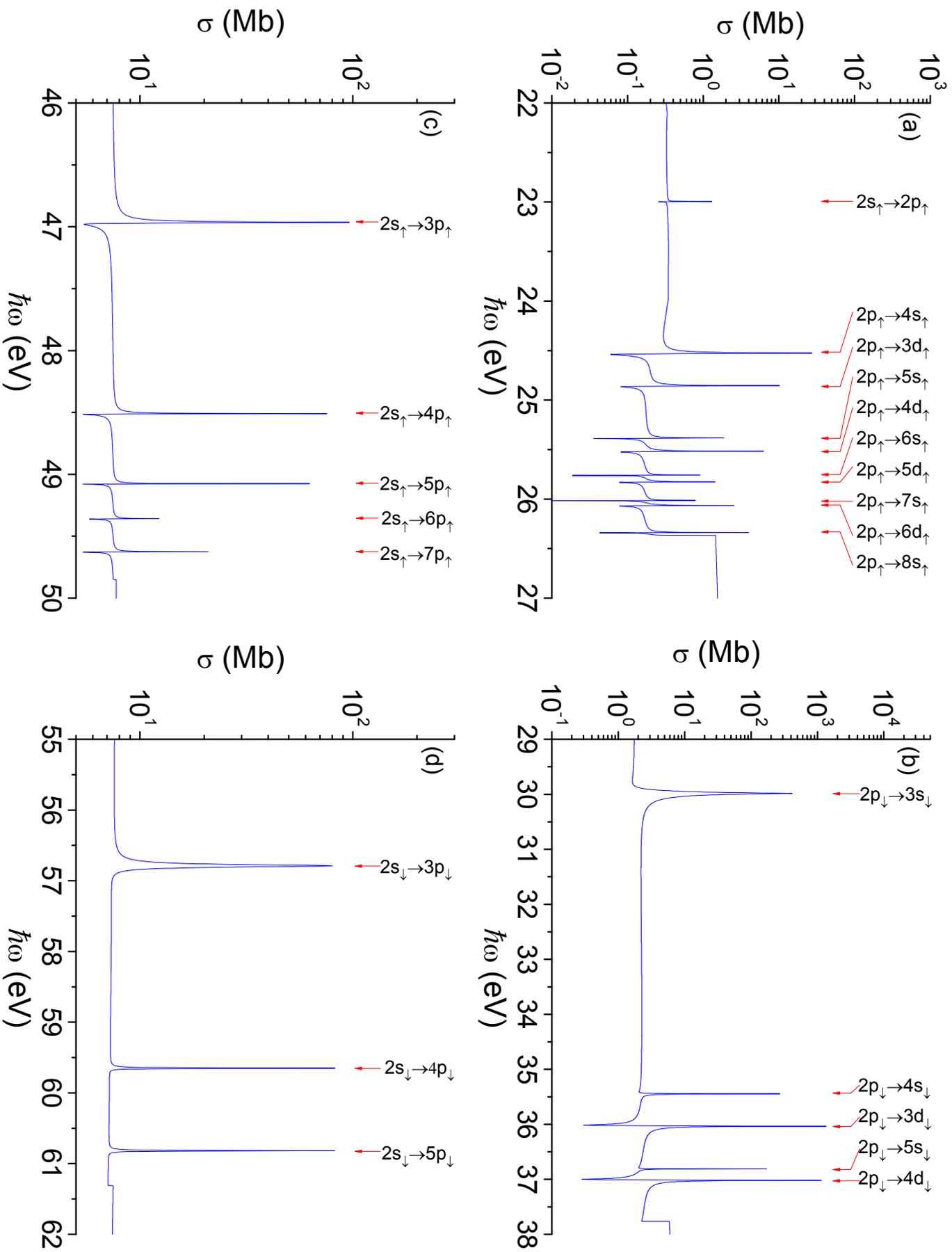
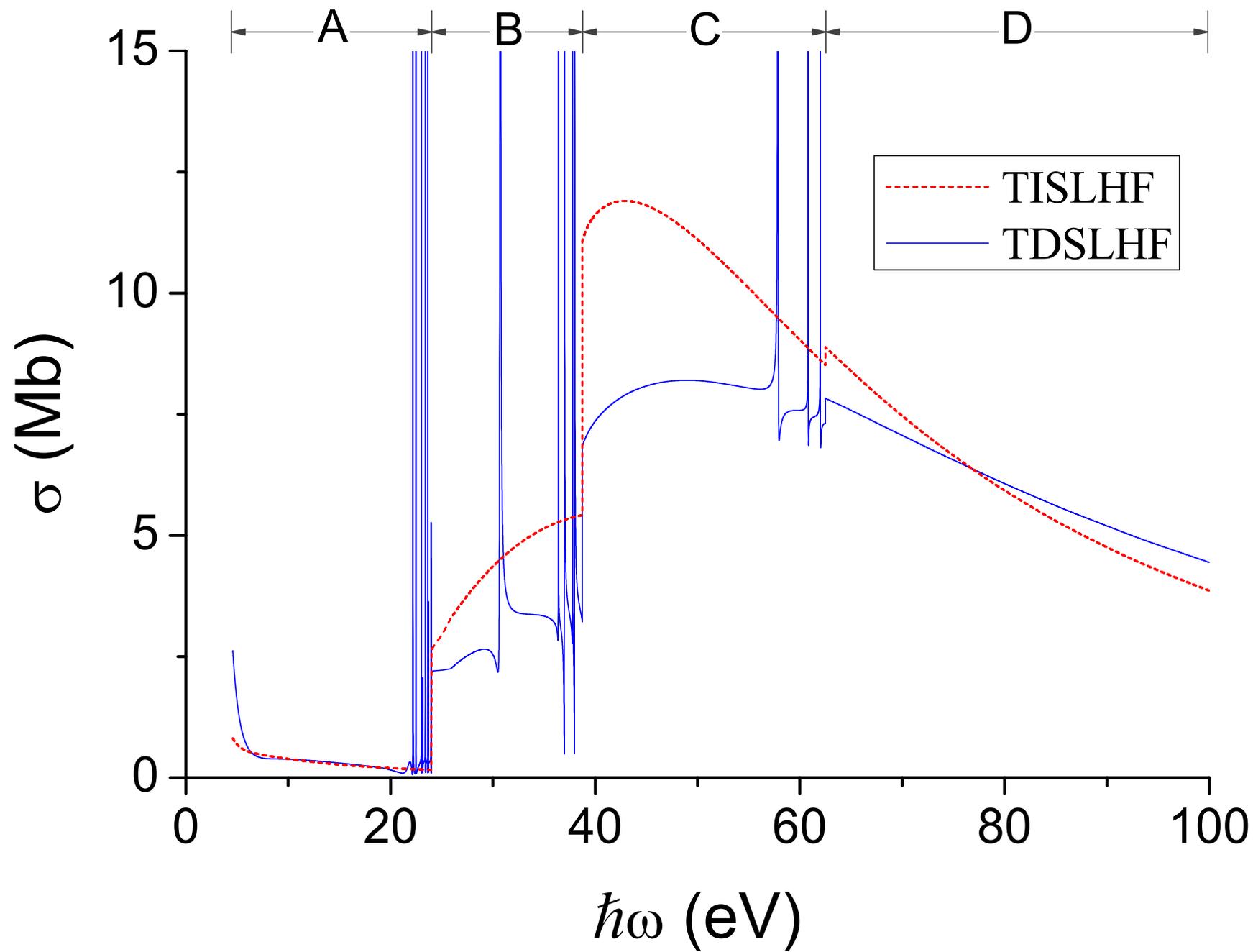
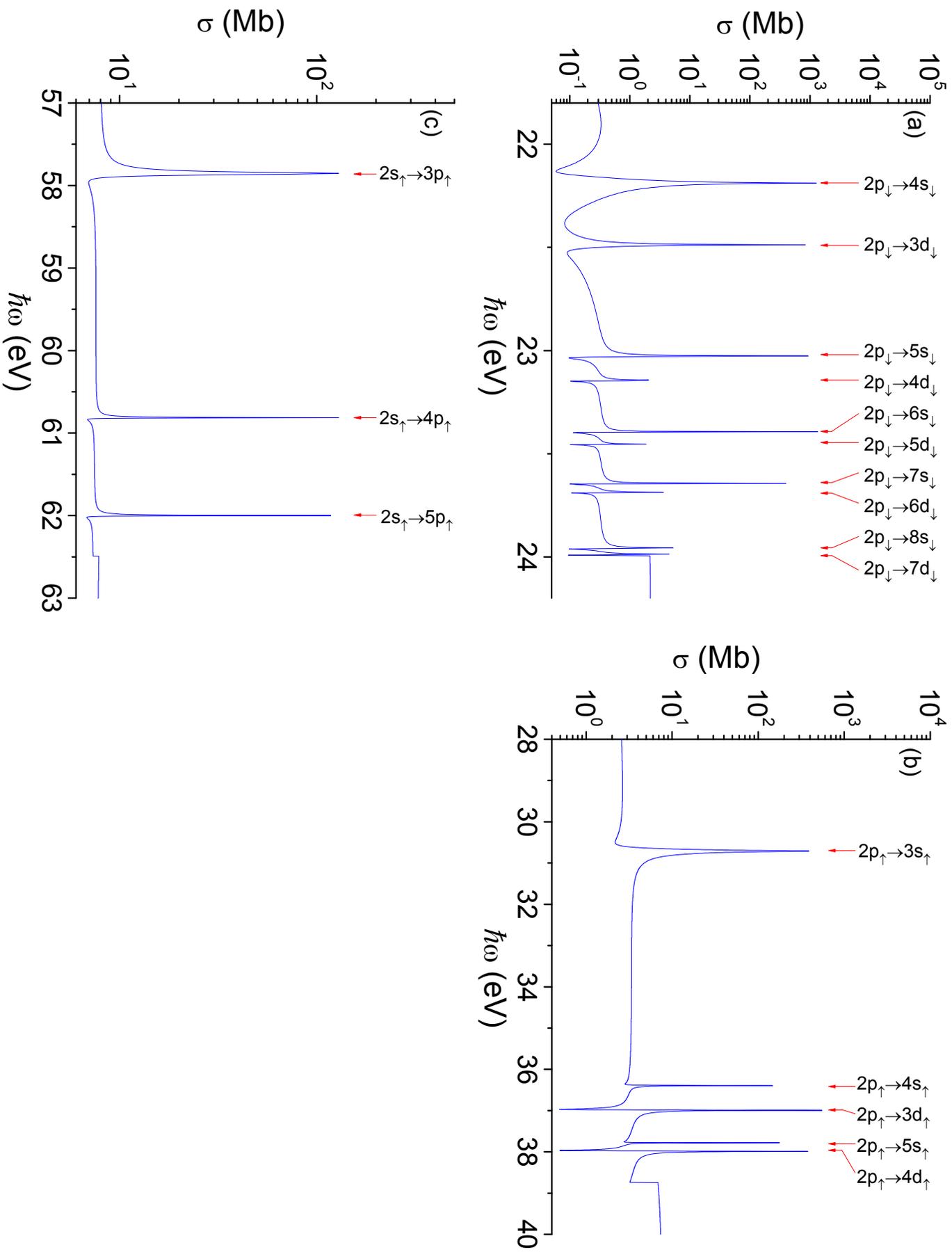
\end{document}